# Transmission of distress in a bank credit network

Yoshiharu Maeno[1], Satoshi Morinaga[1],
Hirokazu Matsushima[2], and Kenichi Amagai[2],

[1] NEC Corporation, 1753 Shimonumabe, Nakahara-ku, Kawasaki,
Kanagawa 211-8666, Japan
[2] Institute for International Socio-Economic Studies, 4-28 Mita, 1-chome, Minato-ku,
Tokyo 108-0073, Japan
y-maeno@aj.jp.nec.com

**Abstract.** The European sovereign debt crisis has impaired many European banks. The distress on the European banks may transmit worldwide, and result in a large-scale knock-on default of financial institutions. This study presents a computer simulation model to analyze the risk of insolvency of banks and the consequent knock-on defaults in a bank credit network. Simulation experiments quantify the worst impact which is imposed on the number of bank defaults by heterogeneity of the bank credit network, the equity capital ratio of banks, and the capital surcharge on big banks.

**Keywords:** Bank credit network, Capital surcharge, Insolvency, Knock-on default.

## 1  Introduction

The European sovereign debt crisis has impaired many European banks. European Banking Authority made a warning announcement that European banks are short of equity capital summing up to EUR 114.7 billion, and presented a recapitalization plan[1] on December 18, 2011. The financial distress on some European banks may transmit to banks across country borders, and causes a knock-on default of banks and financial institutions worldwide. Indeed, many globally operating financial institutions like Lehman Brothers, Merrill Lynch, and American International Group ended in bankruptcy in the financial crisis which ensued from the collapse of the subprime mortgage market in 2007. Since then, supervisors and other relevant authorities have directed a great effort at comprehending the risk which is hidden behind globally interlinked financial institutions and at finding the solution to contain the catastrophic worldwide transmission of distress.

For these purposes, computer simulation models [2] have been developed to understand the contagion in financial stability [5], and financial fragility [6]. The works of particular interest are the studies on the number of bank defaults in a knock-on default [10], systemic risk [1], and the impact that the impairment of a bank imposes on the balance sheet of the other banks [8], and the effect of the liquidity of assets on a knock-on default [7]. These are abstract models, and not founded solidly on the observed structural properties of real bank credit networks. Hence, the models do not provide supervisors and other relevant authorities with any concrete information to aid them in standardizing the best practice as regulatory policies. On the other hand, some recent works include the studies on the detailed process where a bank go bankrupt [4], on the structural properties of the funds transfer between banks [11], and on the business loans from banks to companies [12]. It is, therefore, a technical issue to incorporate the knowledge which is learned from these works in the computer simulation models.

---
[1] A. Enria, Results of bank recapitalization plan,
  http://eba.europa.eu/capitalexercise/2011/2011-EU-Capital-Exercise.aspx.

In this study, a computer simulation model is developed to analyze the risk of insolvency of banks and the consequent knock-on defaults in a bank credit network. In the model, the topology of a bank credit network represents interbank loans, two quantities represent the assets in the balance sheet of individual banks, and three quantities represent the liabilities. This study does not address the liquidity risk but the risk of insolvency. The model does not include any roles of clearing houses either. Simulation experiments with the model quantify the worst impact which is imposed on the number of bank defaults, as a final outcome of the knock-on default, by heterogeneity of the bank credit network, the equity capital ratio of banks, and the capital surcharge on big banks.

## 2 Model

### 2.1 Balance sheet

An interbank loan is the credit relation between a creditor bank and a debtor bank which appears when the debtor bank raises money in the interbank market. A bank credit network describes the all credit relations. It is a directed graph which consists of banks as vertices, and the loans as edges from creditor banks to debtor banks. The amount of the interbank loan and interbank borrowing of individual banks is determined consistently by the given topology of the bank credit network [1]. The balance sheet of individual banks is determined as follows. The balance sheet represents the financial state of a bank at a moment. Figure 1 shows the model for the balance sheet of a bank.

$A_i$ is the asset of the $i$-th bank ($i = 1, \ldots, N$). The number of banks in the bank credit network is $N$. $A_i$ consists of an external asset $E_i$ and interbank loans $I_i$. $I_i$ is the interbank borrowings of other banks. $A_i=E_i+I_i$ holds. The external asset is an investment in general, for example, financing to companies and investing in securities. It is assumed in this study that the price of the external asset does not change and that the insolvency of a debtor bank for the interbank borrowing causes distress to the creditor bank and may ignite a knock-on default from the debtor bank.

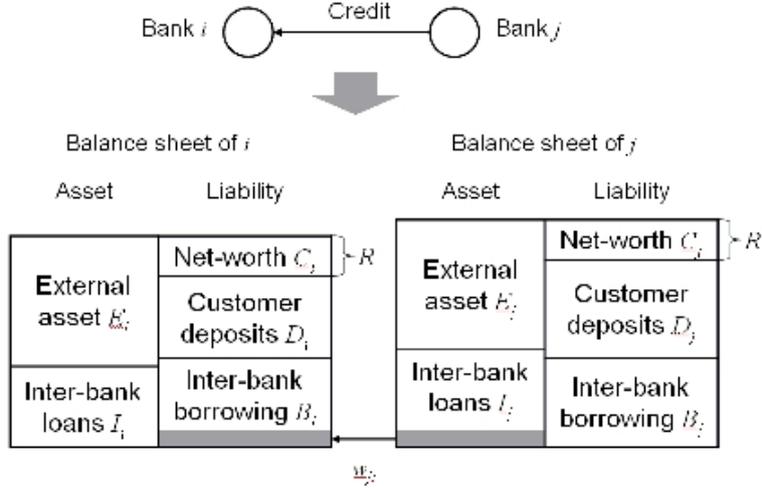

**Fig. 1.** Interbank borrowing of a debtor bank from a creditor bank, and the resulting balance sheet of the two banks.

$L_i$ is the liability of the $i$-th bank. $L_i$ consists of the net worth $C_i$, the interbank borrowing $B_i$, and the customer deposits $D_i$. $B_i$ is the interbank loans of other banks. The net worth means the equity capital which represents the core tier 1 capital including common stock and disclosed reserves. These need not be paid off, and can be used to absorb the loss from the distress immediately. $L_i=A_i$ and $L_i=C_i+B_i+D_i$ hold. Four out of these five quantities are independent variables.

Two constants $Q$ and $R$ describe the characteristics of the bank credit network. The constant $Q$ is the total interbank loans $I=\Sigma I_i$ as a fraction of the total assets $A=\Sigma A_i$. Give the amount of the total external assets $E=\Sigma E_i$, Equation (1) holds.

$$I = QA = \frac{Q}{1-Q}E \quad (1).$$

The constant $R$ is the equity capital ratio of banks. The equity capital ratio of individual bank is defined by Equation (2).

$$R_i = \frac{C_i}{L_i} \quad (2).$$

It is assumed in this study that both big banks and small banks have the same value of $R$. The value of $R$ is the minimal level of the equity capital ratio that is required by the bank regulatory policies.

The balance sheet of individual banks is determined consistently, given the topology of the bank credit network, and the constants $Q$ and $R$. A matrix $l$ represents the topology. If the $j$-th bank borrows money from the $i$-th bank, $l_{ij}=1$, and $l_{ij}=0$ otherwise. In general, $l_{ij}=l_{ji}$ does not hold. The number of the debtor banks which borrow from the $i$-th bank is $g_i$. This is the number of outgoing edges of the vertex. The number of the creditor banks from which the $i$-th bank borrows is $c_i$. This is the number of incoming edges. These are called nodal degrees. The number of edges as a fraction of the number of the pairs between banks is $p$. In other words, it is the number of elements in $l$ whose value is $l_{ij}=1$ as a fraction of $N(N-1)$. The average of the nodal degrees satisfies Equation (3).

$$p = \frac{\overline{g_i}}{N-1} = \frac{\overline{c_i}}{N-1} \quad (3).$$

The amount of the interbank borrowing of the $j$-th bank from the $i$-th bank is $w_{ij}$ given by Equation (4).

$$w_{ij} = \frac{l_{ij}g_i^s c_j^t}{\sum_{j \neq i} l_{ij}g_i^s c_j^t} \frac{Q}{1-Q} E \quad (4).$$

The value of these two powers, $s$ and $t$, characterize the heterogeneity in the amount of the loans. As the value of the powers increases, larger fraction of loans is owned by big banks which usually have large nodal degrees. The heterogeneity of the interbank loans emerges. If $s=0$ and $t=0$, the amount of loans is the same among any pairs of banks. This was the assumption made in most of the previous studies.

The amount of the interbank loan and interbank borrowing is the sum of $w_{ij}$, and is given by Equations (5) and (6).

$$I_i = \sum_{j \neq i} w_{ij} \quad (5).$$

$$B_i = \sum_{j \neq i} w_{ji} \quad (6).$$

A prerequisite $E_i > B_i - I_i$ is imposed on the balance sheet of individual banks. That is, the external asset is no less than the net interbank borrowing. Otherwise, those banks have already gone bankrupt. $E_i$ is given by Equation (7).

$$E_i = \max(B_i - I_i, 0) + \frac{E - \sum_{i=1}^{N} \max(B_i - I_i, 0)}{N} \quad (7).$$

The prerequisite is satisfied because of the first term on the right side of Equation (7). The second term, which is the same among banks, is added to the first term. The values of $A_i$, $L_i$, $C_i$, and $D_i$ are determined from these values of $I_i$, $B_i$, and $E_i$.

## 2.2 Default condition

Given the balance sheet of banks, a simulation experiment reproduces a knock-on default which ensues from the bankruptcy of a particular bank. The distress whose strength is $S_i = E_i$ strikes an arbitrarily chosen bank initially. If the equity capital can absorb the distress ($C_i > S_i$), the bank does not go bankrupt, nor the distress transmit. Otherwise, the $i$-th bank goes bankrupt. The distress transmits to the creditor banks from which the $i$-th bank borrows. The strength of the distress which is transmitted from the $i$-th bank (which went bankrupt) to the $j$-th bank (which is a creditor) is given by Equation (8).

$$S_j = \frac{w_{ji}}{\sum_{j'=1}^{N} w_{j'i}} \max(S_i - C_i, B_i) \quad (8).$$

If the equity capital can absorb the distress ($C_j > S_j$), the $j$-th bank does not go bankrupt. Otherwise, a knock-on default occurs, and the distress transmits to $k$-th bank further. The distress may transmit from multiple banks which go bankrupt to a bank from which those banks borrow. The sum of the distress strikes the bank in this case. The number of bank defaults $N_d$ is the number of defaults in the knock-on default plus one (the initial bankruptcy). Given $Q$ and $p$, the distribution of $N_d(R)$ is investigated as a function of the equity capital ratio $R$ for randomly synthesized 100,000 different topologies of bank credit networks.

Note that this study does not address the liquidity risk that selling assets in the market makes a big loss immediately after some banks go bankrupt, but the risk of the insolvency of a bank for an interbank borrowing because of the shortfall in the capital. The above model does not include any roles of clearing houses either. The clearing houses may reduce the settlement risk by netting loans and borrowings. Netting has the effect similar to delivery-versus-payment transactions like the simultaneous exchange of the title to an asset and payment.

## 2.3 Network topology

Two categories of the topologies $l$ of bank credit networks are investigated in this study. Figure 2 shows an example of homogeneous bank credit networks when $N=500$, as a practical number of banks which operates internationally, and $p=0.005$. The topology is Erdos-Renyi model [14]. In graph theory, the Erdos-Renyi model is a random graph which sets an edge between every pair of vertices with equal probability $p$, independently of the other edges. The distribution of the nodal degree $g$ is binomial, $P(g) = {}_{N-1}C_g\, p^g (1-p)^{N-1-g}$. The value of powers are $s=0$ and $t=0$. The size of vertices indicates the amount of the asset of banks, and the thickness of edges indicates the amount of loans. Banks are homogeneous in the amount of assets and loans.

Figure 3 shows an example of heterogeneous bank credit networks when $N=500$ and $p=0.005$. The topology is Barabasi-Albert model [13]. The Barabasi-Albert model is a random graph with the mechanism of growth and preferential attachment, which becomes scale-free as the number of vertices $N$ goes to infinity, that is, the distribution of the nodal degree $g$ obeys the power law, $P(g) = g^{-a}$. There is a significant probability of the presence of very big vertices. This is the origin of heterogeneity. Natural and human-made systems, including the Internet, citation networks, and social networks are known scale-free. The value of powers are $s=2$ and $t=2$. Big banks are much bigger than small banks.

The characteristics of the bank credit network are studied from the data on the funds transfer by Fedwire [11]. Fedwire is an on-line real-time gross settlement funds transfer system in the United States. It is run by Federal Reserve Banks. The characteristics found from the Fedwire data are summarized below.

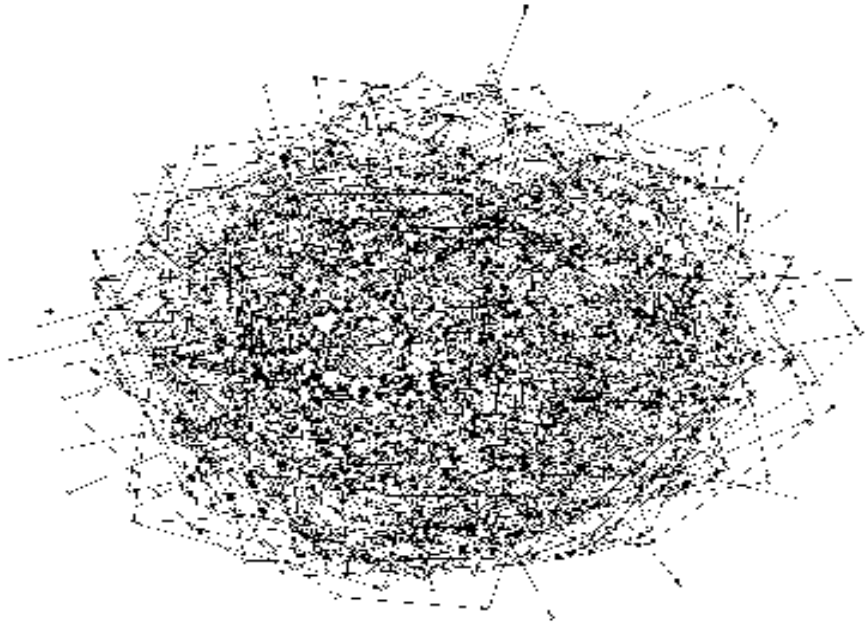

**Fig. 2.** Homegeneous bank credit network when $N$=500 and $p$=0.005.

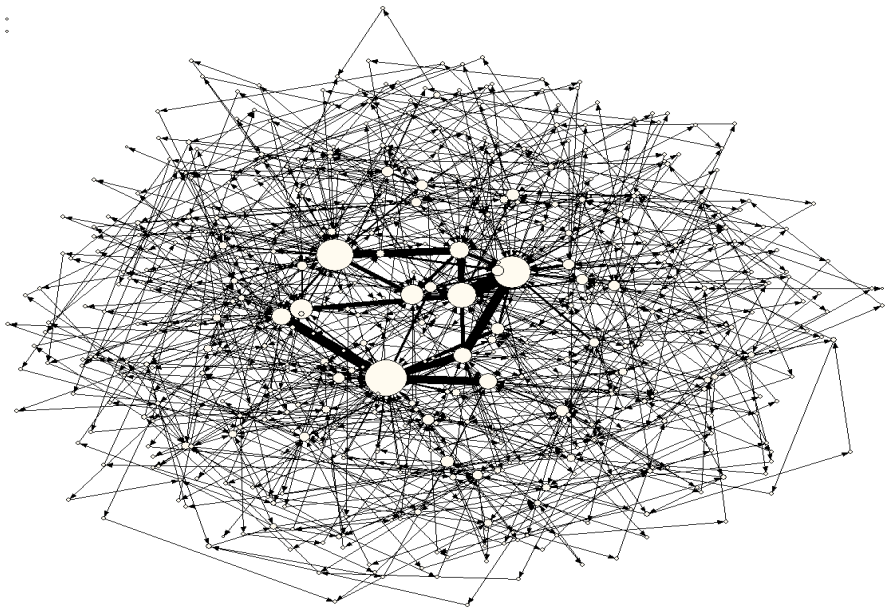

**Fig. 3.** Heterogeneous bank credit network when $N$=500 and $p$=0.005.

- The network consists of 6,600 commercial banks as vertices, 70,000 edges between them, and a clique of 25 big banks as a core sub-structure. The clique is a sub-network where every vertex has edges to the rest.
- Banks are heterogeneous in their size. A few big banks transfer funds to more than 1,000 destination banks while a number of small banks transfer funds merely to a few destination banks.
- The number of pairs of banks where funds are transferred is smaller than 0.5% of the number of possible pairs of banks.
- The distribution of the nodal degree of vertices g and c obeys the power law, $P(g) = g^{-2}$ and $P(c) = c^{-2}$. The amount of funds transferred along an edge increases as the nodal degree at its ends increases.
- There are significant amount of funds which are transferred between big banks and small banks.
- The amount of funds transferred from a bank to a destination bank increases as the number of destination banks to which the bank transfers funds increases. Just 1% of banks transfer funds as much as 75% of the total funds.

The characteristics of the network in Figure 3 are nearly equal to those of Fedwire.

## 3 Default risk

### 3.2 Number of defaults

Figure 4 shows the number of defaults $N_d(R)$ in a knock-on default as a function of the equity capital ratio $R$ for homogeneous bank credit networks when $N=500$, $Q=0.1$, $p=0.005$. The curves in the figure show the 99th percentile, the 95th percentile, the 90th percentile, the mean, and the mean plus one standard deviation. The 99th percentile point represents the worst case. The shape of these curves is similar. The number of defaults decreases largely as $R$ increases. The knock-on default disappears when $R > 0.05$. The mean does not change much at $N_d=3$ when $0.02 < R < 0.04$. Under this condition, the knock-on default transmits to the neighbor banks (the first hop banks) of the bank which went bankrupt initially, but not to the second hop banks of those first hop banks. The average number of the neighbor banks is $pN = 2.5$.

Figure 5 shows $N_d(R)$ as a function of $R$ for heterogeneous bank credit networks when $N=500$, $Q=0.1$, $p=0.005$. The shapes of the curves are different between figures 4 and 5. The effect of increasing the equity capital ratio is much smaller in Figure 5 than that in Figure 4 while $N_d$ still decreases as $R$ increases. When $R$ is relatively large, the effect of increasing the equity capital ratio diminishes. For example, the number of defaults at the 99-th percentile does not decrease much even as $R$ becomes larger than 0.07. The equity capital ratio of $R > 0.15$ is necessary to contain the knock-on default perfectly. The gap between the curves of the 99th percentile and 95th percentile is large. It means that infrequent but catastrophic crisis may happen. This is a warning that a knock-on default may cause a loss which is significantly larger that that predicted from the mean in a real bank credit network like Fedwire.

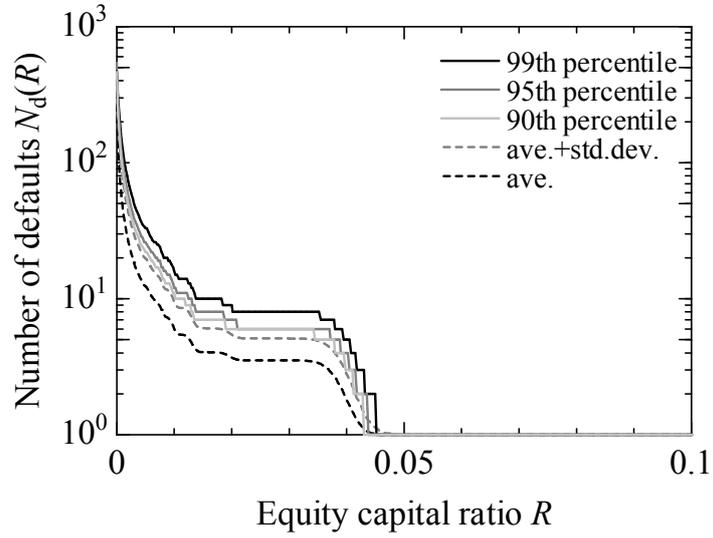

**Fig. 4.** Number of defaults $N_d(R)$ as a function of the equity capital ratio $R$ for homogeneous bank credit networks when $N=500$, $Q=0.1$, and $p=0.005$.

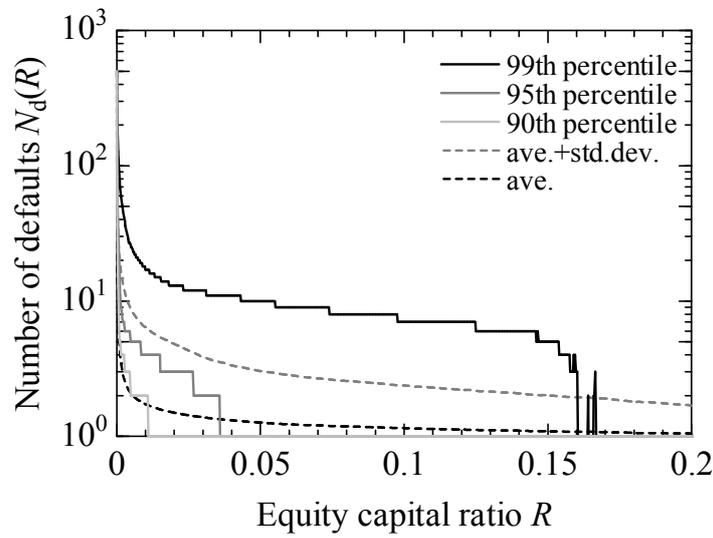

**Fig. 5.** Number of defaults $N_d(R)$ as a function of the equity capital ratio $R$ for heterogeneous bank credit networks when $N=500$, $Q=0.1$, and $p=0.005$.

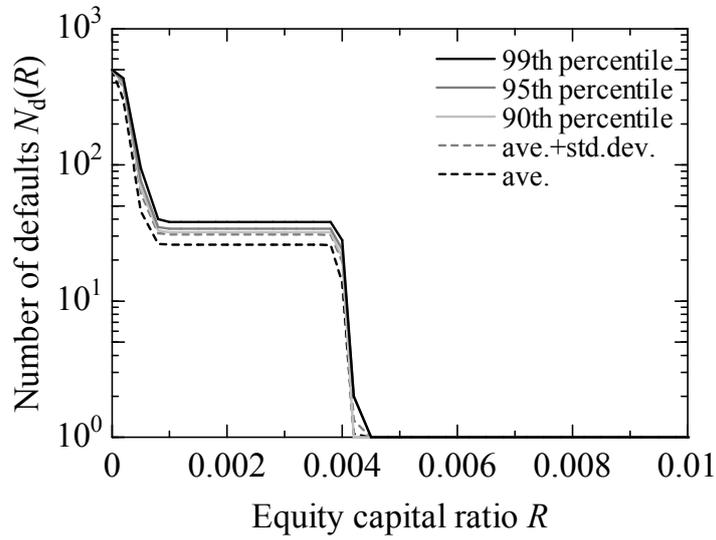

**Fig. 6.** Number of defaults $N_d(R)$ as a function of the equity capital ratio $R$ for homogeneous bank credit networks when $N=500$, $Q=0.1$, and $p=0.05$.

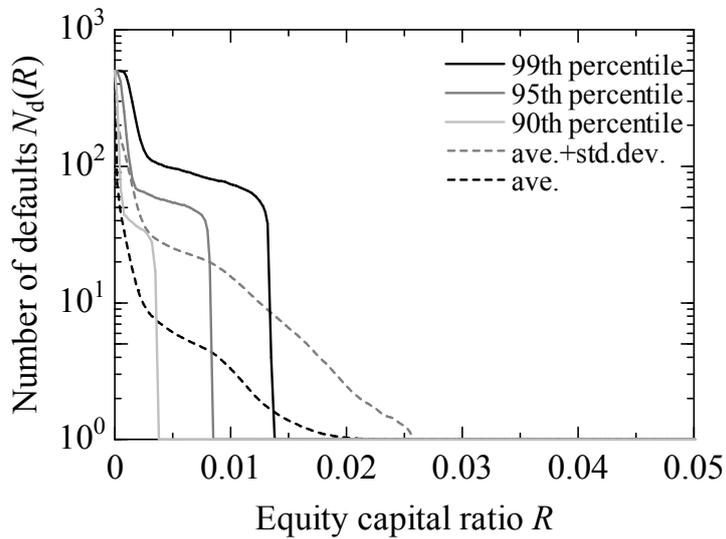

**Fig. 7.** Number of defaults $N_d(R)$ as a function of the equity capital ratio $R$ for heterogeneous bank credit networks when $N=500$, $Q=0.1$, and $p=0.05$.

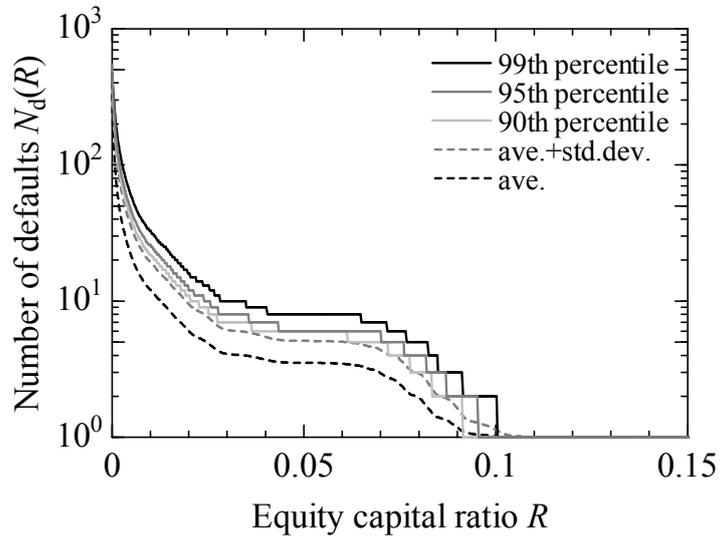

**Fig. 8.** Number of defaults $N_d(R)$ as a function of the equity capital ratio $R$ for homogeneous bank credit networks when $N=500$, $Q=0.2$, and $p=0.005$.

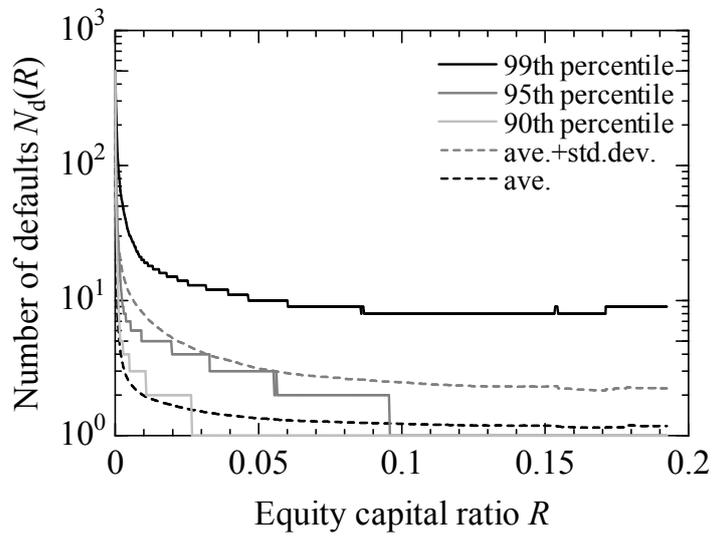

**Fig. 9.** Number of defaults $N_d(R)$ as a function of the equity capital ratio $R$ for heterogeneous bank credit networks when $N=500$, $Q=0.2$, and $p=0.005$.

Figure 6 shows the number of defaults $N_d(R)$ for homogeneous bank credit networks when $N=500$, $Q=0.1$, $p=0.05$. Figure 7 shows $N_d(R)$ for heterogeneous bank credit networks. Knock-on defaults disappear when $R > 0.005$ for homogeneous bank credit networks and $R > 0.03$ for heterogeneous bank credit networks. Dense bank credit networks are more robust. Figure 8 shows the number of defaults $N_d(R)$ for homogeneous bank credit networks when $N=500$, $Q=0.2$, $p=0.005$. Figure 9 shows $N_d(R)$ for heterogeneous bank credit networks. Knock-on defaults disappear when $R > 0.1$ for homogeneous bank credit networks, but they still appear even when $R > 0.2$ for heterogeneous bank credit networks. As the total interbank loans increases, the bank credit networks become less robust.

### 3.2 Effect of surcharge

The Basel Committee on Banking Supervision formulates the standard of the best practice of bank supervision as bank regulatory policies. The policies specify the minimal level of the amount of the equity capital for loss absorbency. The policies have been revised to improve the quality of the equity capital, and to enhance the risk coverage since the 2007 financial crisis. A consultative document[2] set out the proposal from the Financial Stability Board and the Basel Committee on Banking Supervision on the magnitude of additional loss absorbency of the global systemically important banks on July 19, 2011. The additional loss absorbency is called capital surcharge which is 1%, 1.5%, 2%, or 2.5% in the equity capital ratio.

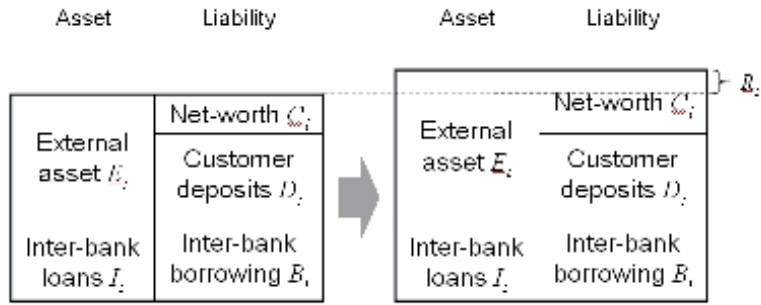

**Fig. 10.** Balance sheet of a big bank on which the additional equity capital ratio $R_s$ is imposed.

The effect of the capital surcharge on big banks on reducing the number of bank defaults is analyzed here. Figure 10 shows the balance sheet of a big bank on which the additional equity capital ratio $R_s$ is imposed. The amount of the net worth becomes $C_i+C'_i$ where $C'_i$ is given by Equation (9). The amount of the asset becomes $A_i+C'_i$.

$$C'_i = \frac{R_s}{1-R-R_s} A_i = \frac{R_s}{1-R-R_s} \frac{C_i}{R} \quad (9).$$

---

[2] Bank for International Settlements, Global systemically important banks: Assessment methodology and the additional loss absorbency requirement, https://www.bis.org/publ/bcbs201.pdf.

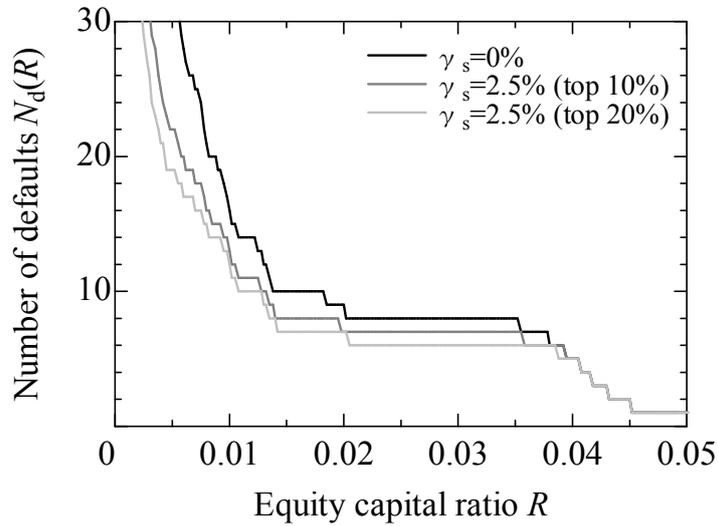

**Fig. 11.** Number of defaults $N_d(R)$ when the additional equity capital ratio $R_s$ is imposed on 0%, 10%, or 20% of the biggest banks in a homogeneous bank credit network. $N$=500, $Q$=0.1, and $p$=0.005.

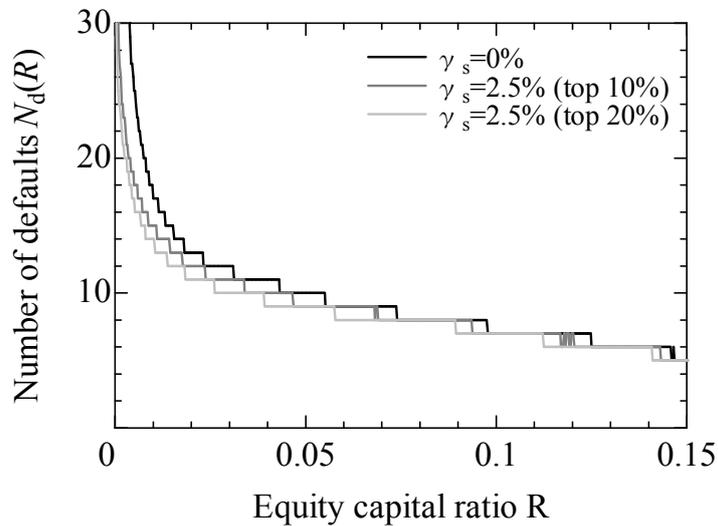

**Fig. 12.** Number of defaults $N_d(R)$ when the additional equity capital ratio $R_s$ is imposed on 0%, 10%, or 20% of the biggest banks in a heterogeneous bank credit network. $N$=500, $Q$=0.1, and $p$=0.005.

Figure 11 shows the number of defaults $N_d(R)$ when the additional equity capital ratio $R_s$=0.025 is imposed on 0%, 10% (50 banks out of $N$=500), or 20% (100 banks) of the biggest banks in a homogeneous bank credit network. The capital surcharge reduces the number of defaults. The effect is stronger if the equity capital ratio is smaller. Figure 12 shows $N_d(R)$ when the additional equity capital ratio $R_s$ is imposed in a heterogeneous bank credit network. The effect of the capital surcharge is not so evident in heterogeneous networks than in homogeneous networks.

No doubt that the capital surcharge strengthens the loss absorbency of individual banks. It reduces the risk that the individual banks go bankrupt, and the burden on tax payers to bail out the banks which are too big to fail. But real bank credit networks like Fedwire are heterogeneous. The capital surcharge may not alleviate the transmission of distress, nor eradicate knock-on defaults. It should not be relied on too heavily to restore the stability of the financial system.

## 4  Conclusion

One of the lessons of the 2007 financial crisis is that supervisors and other relevant authorities failed to sense the risk hidden in the complexity of a heterogeneous bank credit network just by keeping a close eye on individual banks and economic fundamentals. The scope of the argument on the capital surcharge set out in the consultative document is still the resilience of individual global systemically important banks. This study, however, shows that the additional loss absorbency may not work as an efficient fire wall in keeping financial distress from transmitting worldwide completely. The very nature of heterogeneous bank credit networks is a serious hindrance in containing the transmission of distress. Strengthening big banks alone may not solve the problem in restoring the stability of the bank credit network.

The computer simulation models will be extended to address more practical circumstances. Under these, the models will include settlement dates (overnight, short-term, or long-term) of loans, netting of loans, refinancing and other means to raise money, and liquidity of assets and other market mechanisms. Another issue of the models is that the topology of many real bank credit networks is not static, and can not be observed directly either. In the field of social network analysis, the presence or absence of an edge between two persons as vertices is inferred from the observation on their behavior with a statistical analysis [9]. An edge between two populations can also be inferred [3]. Impending phenomena may be predicted from the inferred topology. Similarly, the stability of the financial system may be predicted from both the observed behavior of individual banks and the inferred topology of the bank credit network. This aids the supervisors and other relevant authorities in designing the financial system theoretically, and in making regulatory policies in practice. These are the goal of the emerging field of systems socio-economics.

**Acknowledgments.** We acknowledge Kenji Nishiguchi, Japan Research Institute, and Hidetoshi Tanimura, Ernst&Young ShinNihon LLC, for their advice and discussion.